\documentclass[a4paper,12pt,aps]{revtex4}
\usepackage{amsmath,amssymb,lmodern}
\usepackage{bbold}
\usepackage{bm}
\usepackage{relsize}
\usepackage{braket}
\usepackage{bigints}

\usepackage{graphicx}
\graphicspath{{figures/}}

\newcommand{\average}[1]{\langle{#1}\rangle}

\DeclareMathOperator{\Tr}{Tr}

\renewcommand{\tt}{\textit}

\begin{document}

\title{Thermal behavior, entanglement entropy and parton distributions}
\author{X. Feal}
\author{C. Pajares}
\author{R.A. Vazquez}
\affiliation{ 
Instituto Galego de F\'{\i}sica de Altas Enerx\'{\i}as \& \\
Departamento de F\'{\i}sica de Part\'{\i}culas \\
Universidade de Santiago de Compostela, 15782 Santiago, SPAIN
}

\date{\today}
\begin{abstract}

The apparent thermalization of the particles produced in hadronic collisions
can be obtained by quantum entanglement of the partons of the initial state
once a fast hard collision is produced. The scale of the hard collision is
related to the thermal temperature. As the probability distribution of these
events is of the form $np(n)$, as a consequence, the von Neumann entropy is
larger than in the minimum bias case. The leading contribution to this entropy
comes from the logarithm of the number of partons $n$, all with equal
probability, making maximal the entropy. In addition there is another
contribution related to the width of the parton multiplicity. Asymptotically,
the entanglement entropy becomes the logarithm of $\sqrt{n}$, indicating that
the number of microstates changes with energy from $n$ to $\sqrt{n}$.

\end{abstract}
\maketitle

\section{Introduction}

Recently, it has been emphasized the importance of the entanglement of the
parton wave function of the initial state
\cite{baker2017,kharzeev2017,bergers2018a,bergers2018b,feal2018,muller2017,kovner2018},
showing that the thermalization of the particles produced in collisions of
small systems objects can be achieved by quantum entanglement of the partons
of the initial state.
The apparent thermalization in high energy collisions is achieved during the
rapid quench induced by the hard collision induced by the collision due to the
high degree of entanglement inside the wave functions of the initial
protons. In this way, we expect that the hard scale $T_{\rm h}\sim p_t$ is
related to the thermal component. The thermal component of charged hadron
transverse momentum distributions in $pp$ collisions at $\sqrt{s}=13$ TeV can
be parameterized as \cite{bylinkin2014a,bylinkin2016,bylinkin2014b}
\begin{align}
  \frac{1}{N_{ev}}\frac{1}{2\pi p_t} \frac{d^2N_{ev}}{d\eta dp_t} =
  A_{th}\exp\left(-\frac{m_t}{T_{\rm th}}\right),
\label{thermal_distribution}
\end{align}
and the hard scattering as 
\begin{align}
\frac{1}{N_{ev}}\frac{1}{2\pi p_t} \frac{d^2N_{ev}}{d\eta dp_t} =
A_h\frac{1}{\left(1+\frac{m_t^2}{nT_{\rm h}^2}\right)^n},
\label{hard_distribution}
\end{align}
where $T_{\rm th}$ is the effective temperature and $T_{\rm h}$ can be
considered as a hard temperature which settles the hard scale. The index $n$,
$T_{\rm h}$ and $T_{\rm th}$ were determined from the fit to the experimental
data. One finds
\begin{align}
\frac{T_{\rm h}}{T_{\rm th}}\simeq 4.2,
\label{temperature_ratio}
\end{align}
with $T_{\rm th}\simeq$ 0.17 GeV at $\sqrt{s}$=13 TeV. The ratio between the
hard and soft scales at Eq. \eqref{temperature_ratio} approximately holds for
any centrality and energy in $pp$ collisions as well as PbPb collisions even
if $T_{\rm h}$ and $T_{\rm th}$ have different values for any centrality,
energy and type of collision \cite{feal2018}. This relation between scales
have been also studied in the Higgs boson transverse momentum distribution, in
the case of Higgs boson decay to $\gamma\gamma$ and in the case of Higgs decay
to four leptons \cite{kharzeev2017}. In these two cases the hard scale is
around twenty times larger than in the previous cases but the ratio of
\eqref{temperature_ratio} still holds.

Concerning the index $n$, it was found that it depends on the energy,
centrality and colliding objects, decreasing with multiplicity for not very
high energy density and increasing with multiplicity in the case of PbPb
collisions at $\sqrt{s}$=2.76 TeV \cite{feal2018}. This behavior and the ratio
between $T_{\rm h}$ and $T_{\rm th}$ can be naturally explained as a
consequence of the clustering of color sources (strings) model
\cite{braun2015,armesto1996,diasdedeus2005}. In this approach, $n$ is the
inverse of the normalized fluctuations of the temperature $T_{\rm h}$. As the
multiplicity increases, the number of different clusters increases and thus
the $T_{\rm h}$ fluctuations (to each cluster corresponds a local temperature
$T_{\rm h}$). In this approach the ratio between temperatures has a defined
value $T_{\rm h}/T_{\rm th}=\pi/\sqrt{2}$.

Concerning the entropy, the pure initial parton state $|\psi\rangle$ with
density matrix $\hat{\rho}=|\psi\rangle\langle\psi|$ have zero von Neumann
entropy $S=-\Tr(\hat{\rho}\ln\hat\rho)=0$. If the partons were truly free and
thus incoherent, as it is assumed in the infinite momentum frame, they would
have a non-zero entropy. In a hard collision, characterized by the transverse
momentum $p_t$, is probed only a part of the proton wave function, localized
in a region $H$ within a cone of radius $\sim 1/p_t$ and length $l\sim 1/mx$,
where $m$ is the proton mass and $x$ is the fraction of energy carried by the
hard parton. If we sum over the complementary unobserved
region $S$ we can calculate the reduced density matrix
\begin{align}
\rho_{H} = \Tr_S\rho,
\end{align}
where
\begin{align}
\rho = |\psi_{HS}\rangle\langle \psi_{HS}|,
\end{align}
and the wave function $|\psi_{HS}\rangle$ is the superposition of a suitable
chosen orthonormal set of states $|\psi_n^H\rangle$ and $|\psi_n^{S}\rangle$
localized in the domains $H$ and $S$,
\begin{align}
|\psi_{HS}\rangle = \sum_{n} \alpha_n |\psi_n^H\rangle \psi_n^S\rangle.
\end{align}
As
\begin{align}
\rho_H \equiv \Tr_S\rho = \sum_n \langle\psi_n^S|\psi_{HS}\rangle\langle
\psi_{HS}|\psi_n^S=\sum_n |\alpha_n|^2|\psi_n^H\rangle\langle \psi_n^H|,
\label{reduced_density_matrix}
\end{align}
then the von Neumann entropy of this state is given by
\begin{align}
S=-\Tr(\rho_H\log\rho_H) = -\sum_{n} p_n\log p_n,
\end{align}
with $p_n\equiv |\alpha_n|^2$.

The onset $\tau$ of the hard interaction is given by the hard scale
$\tau\sim 1/p_t$, since $\tau$ is small the quench creates a highly excited
multi-particle state. The produced particles have a thermalized spectrum with
a temperature $T_{\rm th}\simeq 1/2\pi\tau \simeq p_t/2\pi$. In QCD at high
energies, using the Balitsky-Kovchegov (B-K) equation, it has been obtained
\cite{kharzeev2017}
\begin{align}
S=\Delta \log s \simeq \Delta
\log\left(\frac{l}{\epsilon}\right)=\Delta \log\left(\frac{1}{x}\right),
\label{balitsky}
\end{align}
with $\Delta=\bar{\alpha}_s \log(r^2Q_S^2)$,where $r$ is the size of the
dipole, $Q_S$ the saturation momentum and $\epsilon$ the Compton wavelength of
the proton. Expression \eqref{balitsky} is very similar to the result for the
entanglement entropy in $(1+1)$ conformal field theory (CFT)
\cite{holzhey1994,calabrese2006}
\begin{align}
S=\frac{c}{3}\log\left(\frac{l}{\epsilon}\right),
\end{align}
being $c$ the central charge of the CFT, which counts the number of degrees of
freedom.

In this paper we study the behavior of the entanglement entropy with the
hardness of the process, and the role played by the temperature
fluctuations. First, we recover the result for the relation between the
conditioned probability for having at last one hard collision and the full
probability. We will show, using renormalization group arguments, that the
conditioned probability must be a gamma distribution. Finally, we compute the
entanglement entropy. The leading term is the logarithm of the multiplicity
similar to the result expressed by Eq. \eqref{thermal_distribution}. In
addition to that there is a second term which depends on the inverse of the
normalized fluctuations. As the scale $p_t$ of the collision increases, this
second term decreases because the size of the hard region is smaller and thus
the possibility of fluctuations of the hard partons of the wave function.  It
is worth to compare the entanglement entropy, $S^c$, corresponding to the
probability of having at least one hard parton, with the entropy $S$
corresponding to have no constraint. The difference $S^c-S$ decreases with the
scale of hardness, vanishing asymptotically. $S^c-\log\langle n \rangle$ as a
function of the energy density presents a maximum, which corresponds to a
turnover of the behavior of the index $n$ of Eq. \eqref{hard_distribution}
with multiplicity. This behavior is naturally explained in the clustering of
color sources model.

The plan of the paper is as follows. In the next section we obtain the
conditioned probability of having a hard collision. Using the renormalization
group and the scale invariance as an argument we obtain the gamma distribution
as the required probability distribution. In section \ref{sec:section3} we
study the entanglement con Neumann entropy for hard events comparing with the
non-constraint entropy, discussing the differences in connection with the
fluctuations on the number of hard partons and with the clustering of color
sources. Finally, in section \ref{sec:section4} the conclusions are presented.

\section{Conditioned Probability}
\label{sec:section2}
Let us consider the probability $p_n$ of having $n$ partons in a given
collision. It has been shown
\cite{diasdedeus1997a,diasdedeus1997b,diasdedeus1997c,diasdedeus1998,braun1998}
that the conditioned probability $p_n^c$ of having $n$ partons and at least
one giving rise to a hard collision is
\begin{align}
p_n^c = \frac{n}{\langle n \rangle}p_n.
\label{hard_probability}
\end{align}
This equation has been obtained not only for hard events but for events of a
type, denoted by $c$, in which for a result to be considered of the type $c$
is enough to have a single $c$ event in at least one of the elementary
collisions.  Examples of this kind are events without a rapidity gap
(non-diffractive events), hard events, annihilation events in $\bar{p}p$
collisions, events with at least one jet and $W^{\pm}$,$Z^0$ events. Let
$N(n)$ be the number of events with $n$ elementary collisions observed in an
hadronic or nuclear collision, we have
\begin{align}
N(n)\equiv\sum_{i=0}^n {n \choose i}\alpha_c^i(1-\alpha_c)^{n-i}N(n),
\label{binomial}
\end{align}
where $\alpha_c$ is the probability of having an event $c$ in an elementary
collision ($0<\alpha_c<1$). If $\alpha_c$ is small equation \eqref{binomial}
becomes
\begin{align}
N(n)=\alpha_cnN(n)+(1-\alpha_cn)N(n),
\label{rare_norare_decomposition}
\end{align}
where from the definition of a type $c$ event the first term of
\eqref{rare_norare_decomposition} is the number of events $N_c(n)$ where a $c$
occurs,
\begin{align}
N_c(n)=\alpha_cnN(n).
\end{align}
If $N$ is the total number of events we have
\begin{align}
\sum_nN(n)=N,\medspace\medspace\medspace\medspace 
\sum_n n^kN(n)\equiv \langle n^k\rangle, 
\end{align}
and, for the total number of events with $c$ occurring
\begin{align}
\sum_n \alpha_cn N(n)=\alpha_c\langle n\rangle N.
\end{align}
This implies, for the probability distribution of having a $c$ event in $n$
collisions
\begin{align}
p_c(n) = \frac{\alpha_cnN(n)}{\alpha_c \langle n\rangle N}
=\frac{n}{\langle n \rangle}p(n),
\label{rare_probability}
\end{align}
which is of the form of Eq.\eqref{hard_probability}. In this equation $n$ is
the number of elementary collisions (parton-parton or nucleus-nucleus,
depending on the case studied) but Eq.\eqref{rare_probability} has been
applied to the multiplicity particle probability distributions, being $p(n)$
the minimum bias multiplicity distribution. Indeed, the equation
\eqref{hard_probability} was checked in the case of production of $W^{\pm},Z^0$
with data of CDF collaboration at Fermilab \cite{diasdedeus1997b}, for the
production of jet events with UA1 collaboration data at SPS
\cite{diasdedeus1997b}, for the production of Drell-Yan pairs in S-U
collisions with NA38 collaboration data \cite{diasdedeus1997c} and for the
annihilation in $\bar{p}p$ collisions \cite{diasdedeus1997b}. In all cases a
good agreement with the experimental data was obtained. Notice that in
Eq.\eqref{hard_probability} the right hand side is independent of $c$ and only
its shape is determined by the requirement of being of the type $c$. In terms
of cross sections the $c$ events are self-shadowed and their cross section can
be written as a function of only the elementary cross sections of a c-event
\cite{blankenbecler1981,pajares1981}.

This selection procedure of the events satisfying certain $c$ criteria can be
repeatedly applied for subsequent $c$ conditions. For instance from these
events with at least one particle with transverse momentum larger than
$p_{t,1}$ one can further select events with at least one particle with
transverse momentum larger than $p_{t,2}$, $p_{t,2}>p_{t,1}$, and so on (there
are cases in which this multiple selection procedure can not be applied more
than once, like non-diffractive or annihilation events). The corresponding
probability distributions to the repeated selection satisfy
\begin{align}
p(n) \to \frac{n}{\langle n \rangle} p(n)\to 
\frac{n^2}{\langle n^2\rangle}p(n) \to
\cdots \frac{n^k}{\langle n^k \rangle}p(n).
\label{convolution_property}
\end{align}
Notice that
\begin{align}
\langle n\rangle_c = \frac{\langle n^2\rangle}{\langle n \rangle},
\label{nc_to_n_relation}
\end{align}
and 
\begin{align}
\langle n\rangle_c-\langle n\rangle = \frac{\langle n^2\rangle-\langle n
  \rangle^2}{\langle n \rangle} \ge 0.
\end{align}
Transformations of the kind of Eq.\eqref{convolution_property} were studied
long time ago by Jona-Lasinio in connection with the renormalization group in
probability theory \cite{jonalasinio1975}, showing that the only stable
probability distribution under such transformations are the generalized gamma
distributions. The simplest one is the gamma distribution. This transformation
has also been studied in connection with self-similarity condition and the KNO
scaling \cite{braun1998},
\begin{align}
\langle n \rangle p_n = \psi\left(\frac{n}{\langle n \rangle}\right)=
\psi(z).
\label{minimumbias_probability}
\end{align}
For the gamma distribution
\begin{align}
\psi(z)=\frac{\beta^k}{\Gamma(k)}z^{k-1}e^{-\beta z}, 
\medspace\medspace k>1,
\label{gamma_distribution}
\end{align}
we have the normalization condition
\begin{align}
1\equiv \sum_n p_n = \sum_n \frac{1}{\langle n \rangle}
\psi\left(\frac{n}{\langle n \rangle}\right) = \int dz \psi(z) = 1,
\end{align}
and
\begin{align}
1\equiv \sum_n \frac{n}{\langle n \rangle} p_n = \int dz z \psi(z),
\end{align}
which forces $\beta=k$. We will use the gamma distribution in our evaluations. 

The gamma distribution appears in different and related frameworks. It is the
stationary solution of the Fokker-Planck equation associated to the Langevin
equation formulated for the temperature time evolution under a multiplicative
white noise produced by the fast quench of a hard collision at a given $p_t$
scale \cite{feal2018,muller2017,wilk2000,biro2005}.

It appears as well as the size distribution of the clusters of strings formed
in a $pp$, $pA$ or $AA$ collision
\cite{braun2015,armesto1996,diasdedeus2005,braun2000a,braun2000b}. At not very
high energy, only single strings are stretched between the partons of the
colliding objects. These strings can be seen as discs of radius $0.2$ fm in
the transverse plane of the scattering. As the energy or centrality of the
collision increases, the number of strings increases and they start to overlap
forming clusters with different number of strings. As the color field inside
the clusters is larger, these clusters fragment into particles according to
this larger color field and thus larger tensions. Above all critical string
density, a large cluster is formed crossing the collision area. Arguments
based on the renormalization of the color field of the clusters of strings
indicate that the cluster size distribution should be a gamma distribution. By
making a convolution of the gamma distribution with the function $\exp(-x
p_t^2)$, which corresponds to the fragmentation of the cluster of size $x$,
Eq.\eqref{hard_distribution} is obtained for the transverse momentum
distribution. In the same way, making the convolution of the gamma
distribution with a Poisson distribution, which corresponds to the
multiplicity distribution for the fragmentation of a cluster of size $x$ (the
size of the cluster controls the mean value of the Poisson distribution) a
negative binomial distribution for the total multiplicity distribution is
obtained. As the inverse of the parameter $k$ controls the normalized width of
the gamma distribution
\begin{align}
\frac{1}{k}= \frac{\langle z^2\rangle -\langle z\rangle^2}
{\langle z \rangle ^2},
\label{normalized_width}
\end{align}
$k$ decreases with the string density as the number of clusters of different
strings grows and the fluctuations increase. Once the large cluster is
obtained, fluctuations decrease and thus $k$ increases
\cite{braun2015,diasdedeus2005}.

The origin of the non-extensive thermodynamics related to the equation
\eqref{hard_distribution} could be the fractal structure of the
thermodynamical system. In reference \cite{deppman2017} it is shown that such
systems present temperature fluctuations following a gamma distribution. The
repetitive fractal structure has to do with the scale transformations
represented by equation \eqref{convolution_property}.

In terms of the reduced matrix density \eqref{reduced_density_matrix}, the
transformation induced by the repeated selection $p_{t,1}< p_{t,2} < \cdots <
p_{t,j}$ translates into a sum over each time a larger region of soft partons,
modifying the probability $p_n=|\alpha_n|^2$ in the way prescribed by the
chain of equation \eqref{convolution_property}.

\section{Entanglement Entropy}
\label{sec:section3}

We will use \eqref{gamma_distribution} to evaluate the entanglement
entropy. The von Neumann entropy for minimum bias events is
\begin{align}
S=-\sum_{n}p_n\log p_n = -\sum_n \frac{1}{\langle n \rangle
}\psi\left(\frac{n}{\langle n \rangle }\right) \log \left(\frac{1}{\langle n
  \rangle }\psi\left(\frac{n}{\langle n \rangle }\right)\right)=\log \langle n
\rangle -\int_0^\infty dz \psi(z)\log\left(\psi(z)\right),
\end{align}
and the von Neumann entropy for type $c$ events, containing at least one hard
collision,
\begin{align}
S^c&=-\sum_n p_n^c\log p_n^c =
-\sum_n\frac{np_n}{\average{n}}\log\left(\frac{np_n}{\average{n}}\right)=-\sum_n
\frac{n}{\average{n}^2}\psi(z)\log
\left(\frac{n\psi(z)}{\average{n}^2}\right) \nonumber \\
&=-\int_0^\infty dz
z\psi(z)\log\left(\frac{z\psi(z)}{\average{n}}\right)=\log\average{n}-
\int_0^\infty dz z \psi(z)\log\left(z\psi(z)\right).
\label{rare_entropy_definition}
\end{align}
Taking for $\psi(z)$ the gamma distribution, we obtain
\begin{align}
S&=\log\average{n}-\log k +k
+\log\Gamma(k)+\frac{1-k}{\Gamma(k)}\partial_k\Gamma(k)\simeq
\log\average{n}+\frac{1}{2}\left[\frac{k-1}{k}+\log
  \left(\frac{2\pi}{k}\right)\right] \nonumber \\
&\to \log\frac{\langle n \rangle}{\sqrt{k}}=\log\langle n \rangle ^{1/2},
\label{minimumbias_entropy}
\end{align}
and
\begin{align}
S^c&=\log\average{n}+k+\log\Gamma(k)-\frac{k}{\Gamma(k)}\partial_k\Gamma(k)
\simeq \log\average{n}+\frac{1}{2}\left[1+\log
  \left(\frac{2\pi}{k}\right)\right]\nonumber\\
&\to \log\frac{\langle n \rangle}{\sqrt{k}}=\log\langle n \rangle ^{1/2},
\label{rare_entropy}
\end{align}
where the last equality of the above relations hold for large $k$ and
$\Gamma(k)$ is the gamma function. 
We observe that the leading term $\log\average{n}$, as $\average{n}\simeq
s^{\Delta}$ it is similar to the one obtained using the B-K equation. The
difference between both entropies reads
\begin{align}
S^c-S = \log k-\frac{1}{\Gamma(k)}\partial_k \Gamma(k) \simeq \frac{1}{2k}
\end{align}
\begin{figure}[h]
\includegraphics[scale=1]{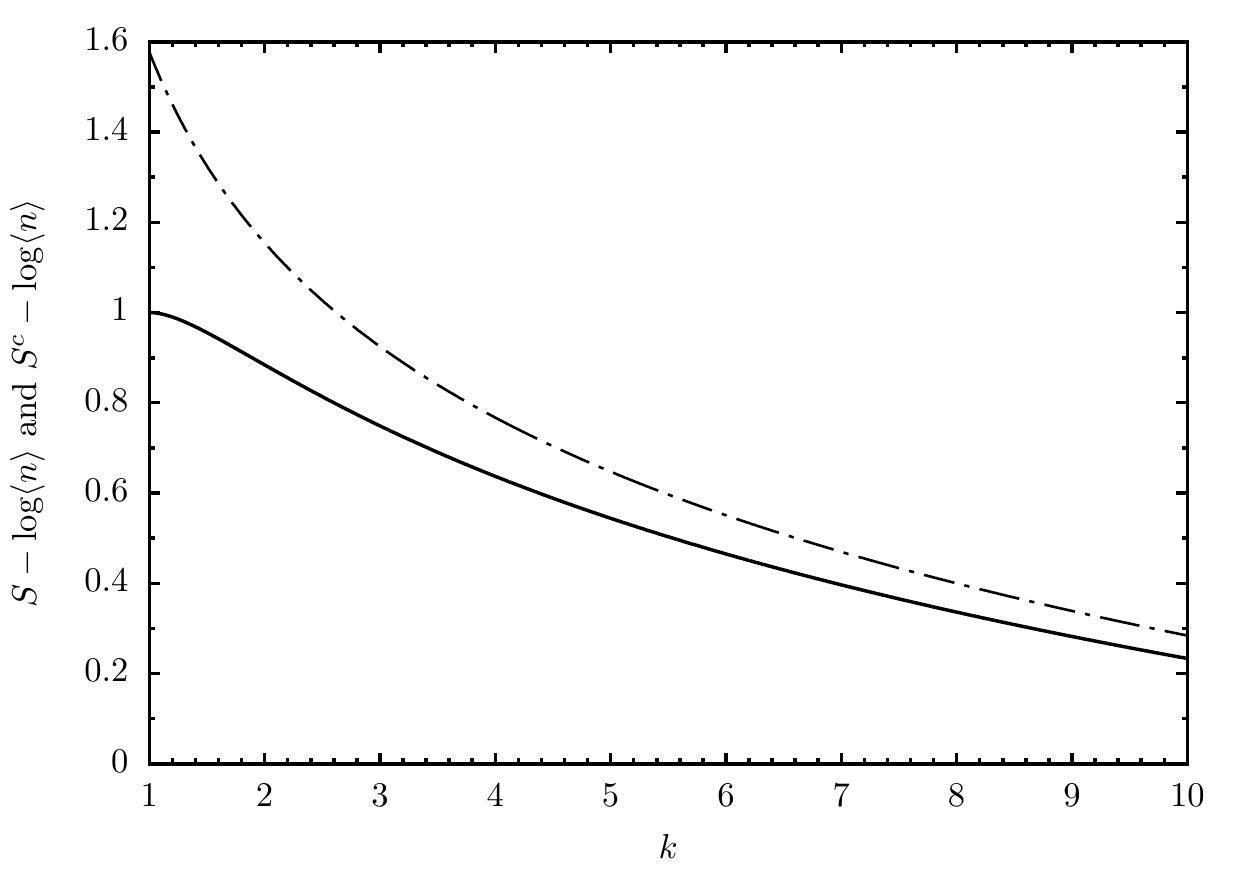}
\caption{Entanglement entropy $S-\log\average{n}$ \eqref{minimumbias_entropy}
  of the minimum bias distribution $p(n)$ \eqref{minimumbias_probability}
  (solid line) and entanglement entropy $S^c-\log\average{n}$
  \eqref{rare_entropy} of the type $c$ events distribution $p_c(n)$
  \eqref{rare_probability} (dot-dashed line).}
\label{fig:figure1}
\end{figure}
In Fig.\eqref{fig:figure1} $S-\log\average{n}$ and $S^c-\log\average{n}$ are
shown as a function of $k$. As $k>1$, $S$ and $S^c$ decrease with $k$ and at
larger values $S^c$ approaches $S$. As $k>1$, $S-\log\langle n \rangle$ and
$S^c-\log\langle n \rangle$ are decreasing functions of $k$ in all the allowed
domain of $k$. These functions, according to Eqs.  \eqref{minimumbias_entropy}
and \eqref{rare_entropy}, become negative at very high $k$. The leading term
of $S$ and $S^c$ is $\log\langle n \rangle$, meaning that the $n$ partons,
i.e. the $n$ microstates of the system, are equally probable and thus the
entropy is maximal. In addition to this contribution, there is one which
depends only on $k$, i.e. the inverse of the normalized fluctuations on the
number of partons, Eq. \eqref{rare_entropy}. This contribution is a positive
decreasing function of $k$ in a very broad range, becoming negative at very
high $k$. In the infinite limit, the gamma function becomes the
normal/Gaussian distribution and both $S$ and $S^c$ behave like $\log(
n/\sqrt{k})=\log( n^{1/2})$. This result means that the number of microstates
is not $n$ any more but $\sqrt{n}$. A saturation effect occurs and the grow of
microstates is suppressed as the collision energy or the centrality increases.
This saturation is explained in models like the color glass condensate or the
clustering of color sources. In this last model, the number of independent
color sources, strings, $n$, formed from the initial partons of the colliding
objects, is reduced at high energies because the number of effective
independent color sources is proportional to $\sqrt{n}$ in such a way that
Eq.\eqref{rare_entropy}, involving logarithms, is satisfied
\cite{diasdedeus2011}. In the limit of high energy in the glasma picture of
the CGC, the number of color flux tubes is also $\sqrt{n}$.

It could be thought that as $\langle n \rangle_c \geq \langle n \rangle$ the
leading term of the entanglement entropy $S^c$ is larger than the
corresponding to $S$. Indeed, instead of Eq.\eqref{rare_entropy_definition} we
could have written
\begin{align}
S^c = \log \average{n}_c -\int dz \psi_c(z)\log\left(\psi_c(z)\right),
\end{align}
with
\begin{align}
\psi_c(z)\equiv \frac{1}{\average{n}_c}p_n^c=
\frac{1}{\average{n}_c}\frac{n}{\average{n}}p_n=\frac{1}{\average{n}_c}\psi(z).
\end{align}
From Eqs.\eqref{normalized_width} and \eqref{nc_to_n_relation} we can write
\begin{align}
\log\average{n}_c=\log\average{n}+\log\left(1+\frac{1}{k}\right),
\label{n_k_relation}
\end{align}
so asymptotically as $k\to\infty$ $\langle n_c\rangle = \langle n \rangle$. 

\begin{figure}[h]
\includegraphics[scale=1]{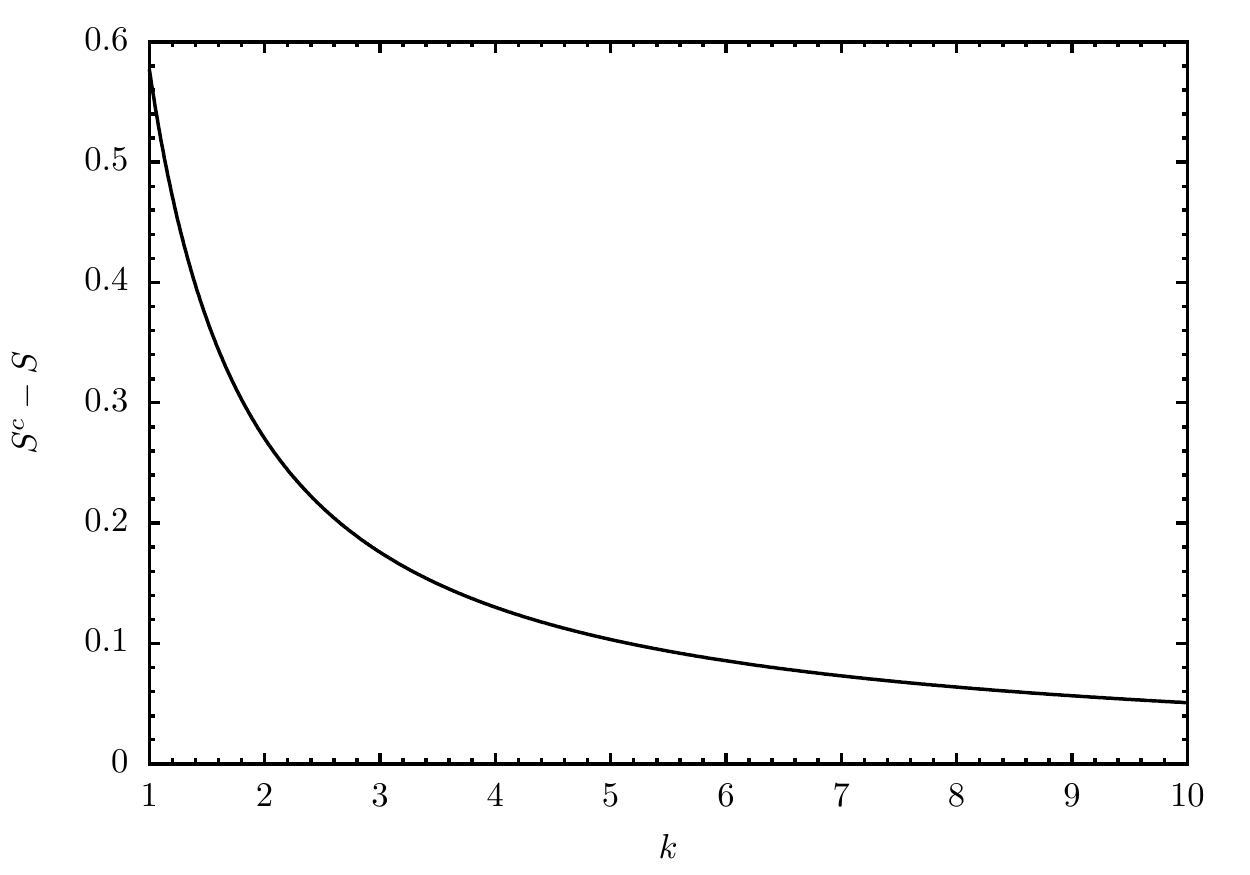}
\caption{Entanglement entropy difference $S^c-S$ at
  \eqref{minimumbias_entropy} and \eqref{rare_entropy} for the minimum bias
  distribution $p(n)$ \eqref{minimumbias_probability} and for the type $c$
  events distribution $p_c(n)$ \eqref{rare_probability}.}
\label{fig:figure2}
\end{figure}

The differences between $S$ and $S^c$ are small and asymptotically tend to
zero as it is shown in Fig.\eqref{fig:figure2}.
The dependence of $S-\log\langle n \rangle$ or $S^c-\log\langle n\rangle$ on
the energy or on the impact parameter (centrality) is very interesting in the
clustering of color sources approach due to the previously described
dependence of $k$ on the string density $\xi$. At low density, $k$ decreases
up to a critical string density $\xi_c$. Above this critical density $\xi >
\xi_c$, $k$ increases. In this way, both $S-\log\langle n \rangle$ and
$S^c-\log\langle n \rangle$ increase with $\xi$ up to the critical density
$\xi_c$, and decrease for larger $\xi$. This decrease of $S-\log\langle n
\rangle$ or $S^c-\log\langle n \rangle$ with energy or centrality is small
compared with the grow of $\log\langle n \rangle$ in such a way that $S$ and
$S^c$ are always growing. The exact value of $k$, which corresponds to the
$\xi_c$ marking the turnover, depends on the observed rapidity range, the
$p_t$ acceptance, and the profile functions of the projectile and target. We
know from the data that $k$ decreases with energy and centrality in pp
collisions and it increases for AuAu and PbPb collisions. The turnover of $k$
could be at very high multiplicity in pp collisions at $\sqrt{s}$ = 13
TeV. Notice that $S$ or $S^c$ does not present a maximum at $\xi=\xi_c$ but a
change in the dependence of $S$ or $S^c$ on $\xi$.

\section{Conclusions}
\label{sec:section4}

Based on the results on the conditioned probability for having at least one
hard collision in terms of the minimum bias probability, we show that this
probability must be the gamma function. This function coincides with the
stationary solution of a Fokker-Planck equation corresponding to a Langevin
equation for the time evolution of the temperature under a multiplicative
white noise produced by the fast quenching of a hard parton. Once the parton
distribution is obtained we compute the entanglement entropy (von Neumann). In
agreement with previous results, the leading term is the logarithm of the
number of partons, meaning that the $n$ microstates are equally probable and
the entropy is maximal. The corrections to the leading term depend only on the
inverse of the normalized fluctuations. Asymptotically, the entanglement
entropy becomes $\log \sqrt{n}$, meaning that the microstates are not anymore
the number $n$ of partons but $\sqrt{n}$ due to saturation effects. We show
that in the clustering of color sources approach $S-\log\langle n\rangle$ or
$S^c-\log\langle n \rangle$ as a function of the energy or of the centrality
should present a maximum corresponding to a critical density of the strings
formed in the collision, which occurs when the overlapping strings cross all
the collision surface.

\section{Acknowledgments}
 We thank the grant Mar\'{\i}a de Maeztu Unit of Excellence of Spain and the
 support of Xunta de Galicia under the project ED431C2017. This paper has been
 partially done under the project FPA2017-83814-P of MCTU (Spain).

\end{document}